\documentclass[journal]{IEEEtran}

\ifCLASSINFOpdf
  \usepackage[pdftex]{graphicx}
  \DeclareGraphicsExtensions{.pdf,.jpeg,.png}
\else
  \usepackage[dvips]{graphicx}
\fi
\usepackage{epstopdf}
\usepackage{amssymb,amsmath,amsthm}
\usepackage{amssymb}
\usepackage{wasysym}
\usepackage{algorithm}
\usepackage{algorithmic}
\usepackage{array}
\usepackage{cite}
\usepackage{color}
\usepackage{url}

\usepackage{epsfig,latexsym}
\usepackage{flushend}
\usepackage{verbatim}
\usepackage{amsopn}
\usepackage{booktabs}

\usepackage{stfloats}
\usepackage{enumerate}
\usepackage{hyperref}
\usepackage{subfigure}
\usepackage{caption}
\usepackage{bm}
\renewcommand{\baselinestretch}{0.93}

\begin{document}

\title{Multi-Beam Forming with Movable-Antenna Array}

\author{{Wenyan Ma, \IEEEmembership{Student Member, IEEE}, Lipeng Zhu, \IEEEmembership{Member, IEEE}, and  Rui Zhang, \IEEEmembership{Fellow, IEEE}}
	 \vspace{-18pt}
\thanks{W. Ma and L. Zhu are with the Department of Electrical and Computer Engineering, National University of Singapore,
	Singapore 117583 (e-mail: {wenyan@u.nus.edu}, {zhulp@nus.edu.sg}).
	
	R. Zhang is with School of Science and Engineering, Shenzhen Research Institute of Big Data, The Chinese University of Hong Kong, Shenzhen, Guangdong 518172, China (e-mail: rzhang@cuhk.edu.cn). He is also with the Department of Electrical and Computer Engineering, National University of Singapore, Singapore 117583 (e-mail: elezhang@nus.edu.sg).
}}
\maketitle

\begin{abstract}
Conventional multi-beam forming with fixed-position antenna (FPA) arrays needs to trade-off between maximizing the beamforming gain over desired directions and minimizing the interference power over undesired directions. In this letter, we study the enhanced multi-beam forming with a linear movable-antenna (MA) array by exploiting the new degrees of freedom (DoFs) via antennas' position optimization. Specifically, we jointly optimize the antenna position vector (APV) and antenna weight vector (AWV) to maximize the minimum beamforming gain over multiple desired directions, subject to a given constraint on the maximum interference power over undesired  directions. We propose an efficient alternating optimization algorithm to find a suboptimal solution by iteratively optimizing one of the APV and AWV with the other being fixed. Numerical results show that the proposed multi-beam forming design with MA arrays can significantly outperform that with the traditional FPA arrays and other benchmark schemes in terms of both beamforming gain and interference suppression.

\end{abstract}

\section{Introduction}
Multi-beam forming is an important signal processing technique in multiple-antenna systems to achieve concurrent signal transmissions to multiple receivers or receptions from multiple transmitters. By designing the amplitude and/or phase of the antenna weight vector (AWV), the transmit/receive signals at different antennas can be constructively superimposed to maximize the beamforming gain over desired directions, or destructively canceled to minimize the interference power over undesired directions. With the development of array signal processing and antenna manufacturing technologies, multi-beam forming has been widely applied e.g., in satellite communications to enhance the coverage of multiple areas \cite{peng2022integrating}, in mobile communications to support broadcasting/multicasting \cite{sidiropoulos2006transmit}, in radar systems to track multiple targets \cite{schulwitz2005compact}, etc. However, since the conventional fixed-position antenna (FPA) arrays usually have a fixed geometry once manufactured, the spatial correlations of steering vectors over different steering angles are fixed for an FPA array. Therefore, existing multi-beam forming methods with the FPA arrays need to trade-off between maximizing the beamforming gain over desired directions and minimizing the interference over undesired directions, which limits the performance of multiple-antenna systems \cite{sidiropoulos2006transmit,karipidis2007far}.

Recently, movable antenna (MA) was proposed as a promising technology for efficiently improving the wireless channel conditions and enhancing the communication performance via the local movement of antennas \cite{zhu2022modeling,ma2022mimo,zhu2023movable,zhu2023movablemagzine,ma2023compressed,chen2023joint,wu2023movable}. Preliminary studies have demonstrated the advantages of MAs in improving the received signal-to-noise ratio (SNR) \cite{zhu2022modeling}, the MIMO channel capacity \cite{ma2022mimo,chen2023joint}, and the performance of multiuser communications \cite{zhu2023movable,wu2023movable} over their FPA counterparts. Moreover, it has been shown in \cite{zhu2023movablebeam} that an MA array can significantly enhance the single-beam forming over FPA arrays by jointly optimizing the antenna position vector (APV) and AWV. Under certain null-steering
directions and numbers of MAs, an MA array can achieve the full beamforming gain over the desired direction with null steering over undesired directions. However, the performance improvement on multi-beam forming with MA arrays has not been investigated in the existing literature \cite{zhu2022modeling,ma2022mimo,zhu2023movable,zhu2023movablemagzine,ma2023compressed,chen2023joint,wu2023movable,zhu2023movablebeam}. 

In this letter, we investigate the enhanced multi-beam forming with a linear MA array by jointly optimizing its APV and AWV. We aim to maximize the minimum beamforming gain over desired directions (i.e., max-min beamforming gain), subject to a given constraint on the maximum interference power over undesired  directions. The formulated optimization problem is non-convex with respect to both the APV and AWV. To address this problem, we propose an alternating optimization algorithm to iteratively optimize one of the APV and AWV with the other being fixed, where the convex relaxation technique is leveraged to find a suboptimal solution for each subproblem. Simulation results show that the proposed multi-beam forming design with MA arrays can significantly outperform that with the traditional FPA arrays and other benchmark schemes in terms of both beamforming gain and interference suppression.

\textit{Notations}: Symbols for vectors (lower case) and matrices (upper case) are in boldface. $\|\bm{a}\|_2$ denotes the $2$-norm of vector $\bm{a}$. We use $(\cdot)^{\mathsf *} $, $(\cdot)^{\mathsf T} $, and $(\cdot)^{\mathsf H} $ to denote conjugate, transpose, and conjugate transpose, respectively. The real part of vector $\bm{a}$ is denoted by ${\rm{Re}}\{\bm{a}\}$. $\mathbb{C}^{P\times{Q}}$ and $\mathbb{R}^{P\times{Q}}$ denote the sets of $P\times{Q}$ dimensional complex and real matrices, respectively. $\bm{a}[p]$ and $\bm{A}[p,q]$ denote the $p$th entry of vector $\bm{a}$ and the entry of matrix $\bm{A}$ in its $p$th row and $q$th column, respectively. ${\rm {Tr}}(\bm{A})$ denotes the trace of matrix $\bm{A}$. $\textrm{diag}(\bm{a})$ denotes a square diagonal matrix with the elements of vector $\bm{a}$ on the main diagonal.

\section{System Model and Problem Formulation}
\begin{figure}[!t]
	\centering
	\includegraphics[width=70mm]{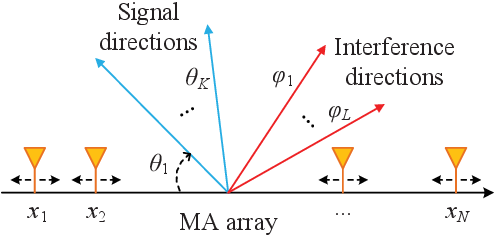}
	\caption{The considered linear MA array.}
	\label{system}
\end{figure}

As shown in Fig.~\ref{system}, we consider a linear MA array with $N$ antennas, where the MAs' positions can be adjusted in the given one-dimensional (1D) line segment of length $D$. Let $x_n\in [0, D]$ denote the $n$th MA's position, and the APV of all $N$ MAs is denoted by $\bm{x}\triangleq[x_1,x_2,\ldots,x_N]^{\mathsf T} \in \mathbb{R}^{N}$ with $0\leq x_1 < x_2 < \ldots < x_N \leq D$ without loss of generality. Thus, the steering vector of the MA array can be written as a function of the APV $\bm{x}$ and the steering angle $\theta$:
\begin{align}
	\bm{\alpha}(\bm{x},\theta) = \left[ e^{j\frac{2\pi}{\lambda}x_1\cos(\theta)}, \ldots, e^{j\frac{2\pi}{\lambda}x_N\cos(\theta)} \right]^{\mathsf T} \in \mathbb{C}^{N},
\end{align} 
where $\lambda$ is the wavelength. Denoting the AWV as $\bm{w} \in \mathbb{C}^{N}$, the beamforming gain over steering angle $\theta$ can thus be expressed as
\begin{align}
	G(\bm{w},\bm{x},\theta)=\left| \bm{w}^{\mathsf H} \bm{\alpha}(\bm{x},\theta) \right|^2.
\end{align}

As shown in Fig.~\ref{system}, let $\{\theta_k\}_{k=1}^K$ denote the set of $K$ desired signals' directions and $\{\phi_l\}_{l=1}^L$ denote the set of $L$ undesired interference directions. Then, we aim to maximize the minimum signal power over $\{\theta_k\}_{k=1}^K$ by jointly optimizing the APV $\bm{x}$ and the AWV $\bm{w}$, subject to a given constraint on the maximum interference power over $\{\phi_l\}_{l=1}^L$, which can be formulated as the following optimization problem:
\begin{subequations}
	\begin{align}
		\textrm {(P1)}~~\max_{\bm{x}, \bm{w}, \delta} \quad & \delta \label{P1a}\\
		\text{s.t.} \quad & x_1\geq 0, x_N\leq D, \label{P1b}\\
		& x_n-x_{n-1} \geq D_0,~~ n = 2,3,\ldots,N,\label{P1c}\\
		& G(\bm{w},\bm{x},\theta_k)\geq \delta,~~ k = 1,2,\ldots,K, \label{P1d}\\
		& G(\bm{w},\bm{x},\phi_l)\leq \eta,~~ l = 1,2,\ldots,L, \label{P1e}\\
		& \|\bm{w}\|_2\leq 1, \label{P1f}
	\end{align}
\end{subequations}
where constraint \eqref{P1b} ensures that the MAs are moved within the feasible region $[0, D]$; $D_0$ in constraint \eqref{P1c} is the minimum distance between adjacent MAs to avoid antenna coupling; $\eta$ in constraint \eqref{P1e} is a given threshold on the maximum interference power; and the AWV power is normalized to be no larger than one in constraint \eqref{P1f}. Note that problem (P1) is a non-convex optimization problem since constraints \eqref{P1d} and \eqref{P1e} are non-convex with respect to $\bm{w}$ and $\bm{x}$. Furthermore, the coupling between $\bm{w}$ and $\bm{x}$ in constraints \eqref{P1d} and \eqref{P1e} renders (P1) more challenging to solve.

\section{Proposed Algorithm}
In this section, we propose an alternating optimization algorithm to tackle the coupling between optimization variables $\bm{w}$ and $\bm{x}$ in problem (P1). Specifically, we iteratively solve two subproblems of (P1) in an alternate manner, where each subproblem optimizes one of $\bm{w}$ and $\bm{x}$ with the other being fixed. Then, the overall algorithm is summarized, and the convergence and computational complexity of the proposed algorithm is analyzed. Finally, a low-complexity initialization scheme is presented.

\subsection{Optimization of $\bm{w}$ with Given $\bm{x}$}
In this subsection, we aim to optimize the AWV $\bm{w}$ in problem (P1) with a given APV $\bm{x}$. Since constraint \eqref{P1d} is non-convex with respect to $\bm{w}$, we use the successive convex approximation (SCA) technique to relax it. Specifically, for the given $\bm{w}^t\in{\mathbb{C}^{N}}$ obtained in the $t$th iteration of SCA, since $G(\bm{w},\bm{x},\theta_k)$ is convex with respect to $\bm{w}$, we can construct the following linear surrogate function $\bar{G}(\bm{w},\bm{x},\theta_k|\bm{w}^t)$ to globally minorize $G(\bm{w},\bm{x},\theta_k)$ by applying the first-order Taylor expansion at $\bm{w}^t$:
\begin{align}\label{G_bar}
	&G(\bm{w},\bm{x},\theta_k)\geq \bar{G}(\bm{w},\bm{x},\theta_k|\bm{w}^t)\\
	&\triangleq G(\bm{w}^t,\bm{x},\theta_k) + 2{\rm{Re}}\{(\bm{w}^t)^{\mathsf H} \bm{\alpha}(\bm{x},\theta_k)\bm{\alpha}(\bm{x},\theta_k)^{\mathsf H} (\bm{w}-\bm{w}^t)\} \notag\\
	&= 2{\rm{Re}}\{(\bm{w}^t)^{\mathsf H} \bm{\alpha}(\bm{x},\theta_k)\bm{\alpha}(\bm{x},\theta_k)^{\mathsf H} \bm{w}\} - G(\bm{w}^t,\bm{x},\theta_k). \notag
\end{align}
Hereto, for the given $\bm{w}^t\in{\mathbb{C}^{N}}$ in the $t$th iteration, the optimization of $\bm{w}$ can be transformed into

\begin{subequations}
	\begin{align}
		\textrm {(P2)}~~\max_{\bm{w},\delta_{\bm{w}}} \quad & \delta_{\bm{w}} \label{P2a}\\
		\text{s.t.} \quad 
		& \bar{G}(\bm{w},\bm{x},\theta_k|\bm{w}^t)\geq \delta_{\bm{w}},~~ k = 1,2,\ldots,K, \label{P2d}\\
		& \eqref{P1e}, \eqref{P1f}. \notag
	\end{align}
\end{subequations}
Since constraint \eqref{P2d} is linear and constraints \eqref{P1e} and \eqref{P1f} are convex quadratic with respect to $\bm{w}$, problem (P2) is a convex quadratically constrained quadratic program (QCQP) problem and can be efficiently solved by using CVX \cite{grantcvx}.

\subsection{Optimization of $\bm{x}$ with Given $\bm{w}$}
In this subsection, we aim to optimize the APV $\bm{x}$ with a given AWV $\bm{w}$. To deal with the non-convex constraints \eqref{P1d} and \eqref{P1e}, we relax them by using the SCA technique. For ease of exposition, we define
$\vartheta_k\triangleq \frac{2\pi}{\lambda}\cos(\theta_k)$.
Furthermore, the $n$th element of $\bm{w}$ is denoted by $w_n = |w_n|e^{j\angle{w_n}}$, with amplitude $|w_n|$ and phase $\angle{w_n}$. Thus, $G(\bm{w},\bm{x},\theta_k)$ in constraint \eqref{P1d} can be further expressed as
\begin{align}
	&G(\bm{w},\bm{x},\theta_k) = \left| \sum_{n=1}^{N} w_n^{\mathsf *} e^{jx_n\vartheta_k} \right|^2 \\
	&= \sum_{n=1}^{N}\sum_{m=1}^{N} |w_n w_m| \cos\big( \vartheta_k(x_n-x_m)-(\angle{w_n}-\angle{w_m}) \big) \notag\\
	&\triangleq \sum_{n=1}^{N}\sum_{m=1}^{N} |w_n w_m| \cos\left( f_k(x_n,x_m) \right), \notag
\end{align}
where $f_k(x_n,x_m) \triangleq \vartheta_k(x_n-x_m)-(\angle{w_n}-\angle{w_m})$. Since $G(\bm{w},\bm{x},\theta_k)$ is neither convex nor concave over $\bm{x}$, we construct a convex surrogate function to locally approximate $G(\bm{w},\bm{x},\theta_k)$ based on the second-order Taylor expansion. Specifically, for a given $z_0\in{\mathbb{R}}$, the second-order Taylor expansion of $\cos (z)$ is
\begin{align}\label{cosz2}
	\cos (z_0)-\sin (z_0) (z-z_0)-\frac{1}{2}\cos (z_0) (z-z_0)^2.
\end{align}
Since $\cos (z_0)\leq 1$ and $(z-z_0)^2 \geq 0$, we can construct the concave quadratic surrogate function $g(z|z_0)$ to minorize $\cos (z)$ as
\begin{align}
	\cos (z)\geq g(z|z_0)\triangleq\cos (z_0)-\sin (z_0) (z-z_0)-\frac{1}{2} (z-z_0)^2.
\end{align}
Then, for the given $\bm{x}^i\triangleq[x_1^i,x_2^i,\ldots,x_N^i]^{\mathsf T}$ in the $i$th iteration of SCA, by letting $z\leftarrow f_k(x_n,x_m)$ and $z_0\leftarrow f_k(x_n^i,x_m^i)$, the surrogate function that provides a global lower-bound for $G(\bm{w},\bm{x},\theta_k)$ can be constructed as
\begin{align}
	&G(\bm{w},\bm{x},\theta_k) \geq \sum_{n=1}^{N}\sum_{m=1}^{N} |w_n w_m| 
	g(f_k(x_n,x_m)|f_k(x_n^i,x_m^i))\notag\\
	&\triangleq \frac{1}{2} \bm{x}^{\mathsf T} \bm{A}_k \bm{x} + \bm{b}_k^{\mathsf T} \bm{x} + c_k,
\end{align}
where $\bm{A}_k \in{\mathbb{R}^{N \times N}}$, $\bm{b}_k \in{\mathbb{R}^{N}}$, and $c_k \in{\mathbb{R}}$ are given by
\begin{align}\label{abc1}
	\bm{A}_k &\triangleq -2\vartheta_k^2 (\gamma\textrm{diag}(\bar{\bm{w}})-\bar{\bm{w}}\bar{\bm{w}}^{\mathsf T}), \\
	\bm{b}_k[n] &\triangleq 2\vartheta_k^2\sum_{m=1}^{N}|w_n w_m|(x_n^i-x_m^i) \notag\\
	&~~~~ - 2\vartheta_k\sum_{m=1}^{N}|w_n w_m|\sin\left( f_k(x_n^i,x_m^i) \right), \notag\\
	c_k &\triangleq \sum_{n=1}^{N} \sum_{m=1}^{N} |w_n w_m|\cos\left( f_k(x_n^i,x_m^i) \right) \notag\\
	&~~~~ + \vartheta_k\sum_{n=1}^{N}\sum_{m=1}^{N}|w_n w_m|\sin\left( f_k(x_n^i,x_m^i) \right)(x_n^i-x_m^i) \notag\\
	&~~~~ - \frac{1}{2}\vartheta_k^2\sum_{n=1}^{N}\sum_{m=1}^{N}|w_n w_m|(x_n^i-x_m^i)^2, \notag
\end{align}
with $\bar{\bm{w}}\triangleq [|w_1|,|w_2|,\ldots,|w_N|]^{\mathsf T}$ and $\gamma\triangleq\sum_{n=1}^{N}|w_n|$. Thus, constraint \eqref{P1d} can be relaxed as a quadratic constraint (QC):
\begin{align}\label{beamgain1}
	\frac{1}{2} \bm{x}^{\mathsf T} \bm{A}_k \bm{x} + \bm{b}_k^{\mathsf T} \bm{x} + c_k \geq \delta_{\bm{x}},~~ k = 1,2,\ldots,K.
\end{align}
Furthermore, by defining $\textbf{1}_{N}\triangleq [1,1,\ldots,1]^{\mathsf T}\in{\mathbb{R}^{N}}$ and $\bm{D}_{\bm{w}}\triangleq\textrm{diag}([\sqrt{|w_1|},\sqrt{|w_2|},\ldots,\sqrt{|w_N|}]^{\mathsf T})\in{\mathbb{R}^{N \times N}}$, we have $\textrm{diag}(\bar{\bm{w}})=\bm{D}_{\bm{w}}\bm{D}_{\bm{w}}$ and $\bar{\bm{w}}=\bm{D}_{\bm{w}}\bm{D}_{\bm{w}}\textbf{1}_{N}$. Then, $\bm{x}^{\mathsf T} \bm{A}_k \bm{x}$ can be further expressed as
\begin{align}\label{xAx}
	\bm{x}^{\mathsf T} \bm{A}_k \bm{x} &= -2\vartheta_k^2 \left(\gamma\bm{x}^{\mathsf T}\textrm{diag}(\bar{\bm{w}})\bm{x}-\bm{x}^{\mathsf T}\bar{\bm{w}}\bar{\bm{w}}^{\mathsf T}\bm{x}\right) \\
	&= -2\vartheta_k^2 \left(\gamma\|\bm{D}_{\bm{w}}\bm{x}\|_2^2 -\left|\textbf{1}_{N}^{\mathsf T}\bm{D}_{\bm{w}}\bm{D}_{\bm{w}}\bm{x}\right|^2\right) \notag\\
	&\overset{(a)}= -2\vartheta_k^2 \left(\|\bm{D}_{\bm{w}}\textbf{1}_{N}\|_2^2 \|\bm{D}_{\bm{w}}\bm{x}\|_2^2 -\left|\textbf{1}_{N}^{\mathsf T}\bm{D}_{\bm{w}}\bm{D}_{\bm{w}}\bm{x}\right|^2\right) \notag\\
	&\overset{(b)}\leq 0, \notag
\end{align}
where the equality $(a)$ holds since $\gamma=\|\bm{D}_{\bm{w}}\textbf{1}_{N}\|_2^2$; and the inequality $(b)$ holds due to the Cauchy–Schwarz inequality, i.e., $\|\bm{u}\|_2^2 \|\bm{v}\|_2^2 \geq |\bm{u}^{\mathsf T}\bm{v}|^2$ for any two equal-sized vectors $\bm{u}$ and $\bm{v}$. Notice from \eqref{xAx} that $\bm{A}_k$ is a negative semi-definite (NSD) matrix. Thus, constraint \eqref{beamgain1} is a convex QC over $\bm{x}$.

On the other hand, since constraint \eqref{P1e} has a similar structure as constraint \eqref{P1d}, we can relax the non-convex constraint \eqref{P1e} by modifying the procedure of constructing the relaxed convex QC constraint \eqref{beamgain1}. Specifically, since $\cos (z_0)\geq -1$ and $(z-z_0)^2 \geq 0$ in \eqref{cosz2}, we can construct the convex quadratic surrogate function $\tilde{g}(z|z_0)$ to majorize $\cos (z)$ as
\begin{align}
	\cos (z)\leq \tilde{g}(z|z_0)\triangleq\cos (z_0)-\sin (z_0) (z-z_0)+\frac{1}{2} (z-z_0)^2.
\end{align}
Then, the surrogate function that provides a global upper-bound for $G(\bm{w},\bm{x},\phi_l)$ can be constructed as
\begin{align}
	&G(\bm{w},\bm{x},\phi_l) \leq \sum_{n=1}^{N}\sum_{m=1}^{N} |w_n w_m| 
	\tilde{g}(\tilde{f}_l(x_n,x_m)|\tilde{f}_l(x_n^i,x_m^i))\notag\\
	&\triangleq \frac{1}{2} \bm{x}^{\mathsf T} \tilde{\bm{A}}_l \bm{x} + \tilde{\bm{b}}_l^{\mathsf T} \bm{x} + \tilde{c}_l,
\end{align}
where $\tilde{f}_l(x_n,x_m) \triangleq \varphi_l(x_n-x_m)-(\angle{w_n}-\angle{w_m})$ with $\varphi_l\triangleq \frac{2\pi}{\lambda}\cos(\phi_l)$; $\tilde{\bm{A}}_l \in{\mathbb{R}^{N \times N}}$, $\tilde{\bm{b}}_l \in{\mathbb{R}^{N}}$, and $\tilde{c}_l \in{\mathbb{R}}$ are given by
\begin{align}\label{abc2}
	\tilde{\bm{A}}_l &\triangleq 2\varphi_l^2 (\gamma\textrm{diag}(\bar{\bm{w}})-\bar{\bm{w}}\bar{\bm{w}}^{\mathsf T}), \\
	\tilde{\bm{b}}_l[n] &\triangleq -2\varphi_l^2\sum_{m=1}^{N}|w_n w_m|(x_n^i-x_m^i) \notag\\
	&~~~~ - 2\varphi_l\sum_{m=1}^{N}|w_n w_m|\sin\left( \tilde{f}_l(x_n^i,x_m^i) \right), \notag\\
	\tilde{c}_l &\triangleq \sum_{n=1}^{N} \sum_{m=1}^{N} |w_n w_m|\cos\left( \tilde{f}_l(x_n^i,x_m^i) \right) \notag\\
	&~~~~ + \varphi_l\sum_{n=1}^{N}\sum_{m=1}^{N}|w_n w_m|\sin\left( \tilde{f}_l(x_n^i,x_m^i) \right)(x_n^i-x_m^i) \notag\\
	&~~~~ + \frac{1}{2}\varphi_l^2\sum_{n=1}^{N}\sum_{m=1}^{N}|w_n w_m|(x_n^i-x_m^i)^2. \notag
\end{align}
Notice that $\tilde{\bm{A}}_l$ is a positive semi-definite (PSD) matrix, which can be proven similarly by the procedure in \eqref{xAx}. Thus, constraint \eqref{P1e} can be relaxed as a convex QC over $\bm{x}$:
\begin{align}\label{beamgain2}
\frac{1}{2} \bm{x}^{\mathsf T} \tilde{\bm{A}}_l \bm{x} + \tilde{\bm{b}}_l^{\mathsf T} \bm{x} + \tilde{c}_l \leq \eta,~~ l = 1,2,\ldots,L.
\end{align}

Therefore, in the $i$th iteration of SCA, $\bm{x}$ can be optimized by solving the following optimization problem
\begin{subequations}
	\begin{align}
		\textrm {(P3)}~~\max_{\bm{x},\delta_{\bm{x}}} \quad & \delta_{\bm{x}} \label{P3a}\\
		\text{s.t.} \quad & \eqref{P1b}, \eqref{P1c}, \eqref{beamgain1}, \eqref{beamgain2}.\notag
	\end{align}
\end{subequations}
Since constraints \eqref{P1b} and \eqref{P1c} are linear and constraints \eqref{beamgain1} and \eqref{beamgain2} are convex quadratic with respect to $\bm{x}$, problem (P3) is a convex QCQP problem and can be efficiently solved by using CVX \cite{grantcvx}.

\subsection{Overall Algorithm and Performance Analysis}
Based on the solutions for problems (P2) and (P3) presented above, we are ready to summarize the overall alternating optimization algorithm for solving problem (P1). The details of the proposed algorithm are summarized in Algorithm~\ref{alg1}, where $\epsilon$, $\epsilon_{\bm{w}}$, and $\epsilon_{\bm{x}}$ are predefined convergence thresholds for the overall algorithm, iterative optimization of $\bm{w}$, and iterative optimization of $\bm{x}$, respectively. Specifically, from step 4 to step 8, we iteratively solve problem (P2) to optimize the AWV $\bm{w}$ with a given APV $\bm{x}$. Then, from step 9 to step 14, we iteratively optimize $\bm{x}$ by solving problem (P3)  with a given $\bm{w}$. The overall algorithm iteratively solves problems (P2) and (P3) until convergence.

\begin{algorithm}[!t]
	\caption{Alternating Optimization Algorithm for Problem (P1)}
	\label{alg1}
	\begin{algorithmic}[1]
		\STATE \emph{Input:} $N$, $L$, $K$, $\eta$, $\{\theta_k\}_{k=1}^{K}$, $\{\phi_l\}_{l=1}^{L}$, $D$, $D_0$, $\epsilon$, $\epsilon_{\bm{w}}$, $\epsilon_{\bm{x}}$, $\bm{x}^0$, $\bm{w}^0$.
		\STATE Initialization:  $t \leftarrow 0$, $i \leftarrow 0$, $\bm{x} \leftarrow \bm{x}^0$, $\delta \leftarrow 0$.
		
		\WHILE{Increase of $\delta$ is above $\epsilon$}	
		
		\WHILE{Increase of $\delta_{\bm{w}}$ in \eqref{P2a} is above $\epsilon_{\bm{w}}$}
		\STATE Obtain $\bm{w}^{t+1}$ by solving problem (P2).
		\STATE $t \leftarrow t+1$.
		\ENDWHILE
		\STATE $\bm{w} \leftarrow \bm{w}^t$, $t \leftarrow 0$.
		
		\WHILE{Increase of $\delta_{\bm{x}}$ in \eqref{P3a} is above $\epsilon_{\bm{x}}$}
		\STATE Obtain $\{\bm{A}_k, \bm{b}_k, c_k\}_{k=1}^K$ via \eqref{abc1}; obtain $\{\tilde{\bm{A}}_l, \tilde{\bm{b}}_l, \tilde{c}_l\}_{l=1}^L$ via \eqref{abc2}.
		\STATE Obtain $\bm{x}^{i+1}$ by solving problem (P3).
		\STATE $i \leftarrow i+1$.
		\ENDWHILE
		\STATE $\bm{x} \leftarrow \bm{x}^i$, $i \leftarrow 0$.
		\STATE $\delta \leftarrow \min\{G(\bm{w},\bm{x},\theta_k)\}_{k=1}^K$.
		
		\ENDWHILE	
		\STATE \emph{Output:} $\bm{w}$, $\bm{x}$.
	\end{algorithmic}
\end{algorithm}

First, we analyze the convergence of Algorithm~\ref{alg1} as follows. In the $t$th iteration for solving problem (P2), we have the following inequalities for constraint \eqref{P2d}:
\begin{align}\label{convergence}
	G(\bm{w}^t,\bm{x},\theta_k) &\overset{(c_1)}= \bar{G}(\bm{w}^t,\bm{x},\theta_k|\bm{w}^t) \\
	&\overset{(c_2)}\leq \bar{G}(\bm{w}^{t+1},\bm{x},\theta_k|\bm{w}^t) \overset{(c_3)}\leq G(\bm{w}^{t+1},\bm{x},\theta_k),\notag
\end{align}
where the equality $(c_1)$ holds since the first-order Taylor expansion in \eqref{G_bar} is tight at $\bm{w}^t$; the inequality $(c_2)$ is valid since $\bar{G}(\bm{w},\bm{x},\theta_k|\bm{w}^t)$ is maximized in the $t$th iteration, while its equality can be attained by letting  $\bm{w}^{t+1}=\bm{w}^t$; the inequality $(c_3)$ holds because $\bar{G}(\bm{w},\bm{x},\theta_k|\bm{w}^t)$ globally minorizes $G(\bm{w},\bm{x},\theta_k)$. Hence, the sequence $\{G(\bm{w}^t,\bm{x},\theta_k)\}_{t=0}^{\infty}$ is monotonically increasing and can converge to its maximum value. Moreover, since $\delta = \min\{G(\bm{w},\bm{x},\theta_k)\}_{k=1}^K$, the increment of $\delta$ can be guaranteed after solving problem (P2).  Similarly, the monotonic convergence for solving problem (P3) can be guaranteed. Moreover, the alternating optimization of variables $\bm{x}$ and $\bm{w}$ guarantees the non-decreasing objective value of problem (P1) during the outer iterations of Algorithm~\ref{alg1}, which is also upper-bounded by a finite beamforming gain. Therefore, Algorithm~\ref{alg1} is guaranteed to converge.

Next, we analyze the computational complexity of Algorithm~\ref{alg1} as follows. Let $I$, $I_{\bm{w}}$, and $I_{\bm{x}}$ denote the maximum number of iterations of repeating steps 4–15 for alternately solving problems (P2) and (P3), that of repeating steps 5-6 for solving problem (P2), and that of repeating steps 10-12 for solving problem (P3), respectively. The complexity for solving the QCQP problem (P2) or (P3) is in the order of  $\mathcal{O}((K+L+2)^{0.5}N(N^2+K+L)\ln(1/\beta))$ with accuracy $\beta$ for the interior-point method \cite{fu2021reconf}. Therefore, the total algorithm complexity is in the order of  $\mathcal{O}((K+L+2)^{0.5}N(N^2+K+L)\ln(1/\beta)(I_{\bm{w}}+I_{\bm{x}})I)$.

\subsection{Initialization}
In this subsection, we propose a low-complexity initialization scheme to obtain $\bm{x}^0$ and $\bm{w}^0$ in Algorithm~\ref{alg1}. Intuitively, all MAs should be sufficiently separated to reduce their coupling effect. Thus, we set $\bm{x}^0$ as uniformly sampled in $[0,D]$, with spacing $\frac{D}{N+1}$, i.e., $\bm{x}^0=[\frac{D}{N+1}, \frac{2D}{N+1}, \ldots, \frac{ND}{N+1}]^{\mathsf T}$. For a uniform linear array (ULA) with APV $\bm{x}^0$, we can obtain the globally optimal AWV $\bm{w}^0$ \cite{karipidis2007far}. Specifically, we first solve the following convex semi-definite programming (SDP) problem to obtain a PSD matrix $\bm{W}^0 \in{\mathbb{C}^{N \times N}}$:
\begin{align}
	\textrm {(P0)}~~\max_{\bm{W}^0, \delta_0} \quad & \delta_0 \\
	\text{s.t.} \quad & {\rm {Tr}}(\bm{W}^0 \bm{\Gamma}(\bm{x}^0,\theta_k))\geq \delta_0,~~ k = 1,2,\ldots,K, \notag\\
	& {\rm {Tr}}(\bm{W}^0 \bm{\Gamma}(\bm{x}^0,\phi_l))\leq \eta,~~ l = 1,2,\ldots,L, \notag\\
	& {\rm {Tr}}(\bm{W}^0)\leq 1, \notag
\end{align}
where $\bm{\Gamma}(\bm{x},\theta) \triangleq \bm{\alpha}(\bm{x},\theta) \bm{\alpha}(\bm{x},\theta)^{\mathsf H}\in{\mathbb{C}^{N \times N}}$. Next, we can obtain $\bm{r}_{\bm{w}}\in{\mathbb{C}^{2N-1}}$ based on $\bm{W}^0$ by
\begin{align}
	\bm{r}_{\bm{w}}[m]=\sum_{p=\max(m-N+1,1)}^{\min(m,N)}\bm{W}^0[p,p-m+N],
\end{align}
for $m = 1,2,\ldots,2N-1$. Indeed, $\bm{r}_{\bm{w}}$ is the autocorrelation vector of $\bm{w}^0$, with $\bm{r}_{\bm{w}}[m]\triangleq\sum_{p=\max(m-N+1,1)}^{\min(m,N)}\bm{w}^0[p] \bm{w}^0[p-m+N]^{\mathsf *}$, $m=1,2,\ldots,2N-1$. Finally, we can reconstruct $\bm{w}^0$ from $\bm{r}_{\bm{w}}$ according to the spectral factorization theorem \cite{karipidis2007far}.

\section{Numerical Results}
In this section, we provide numerical results to validate our proposed design for multi-beam forming with MA arrays. We set the convergence threshold as $\epsilon=10^{-4}$ and $\epsilon_{\bm{w}}=\epsilon_{\bm{x}}=10^{-2}$. The minimum distance between adjacent MAs is set as $D_0=\lambda/2$. The threshold on the maximum interference power is set as $\eta=0.1$.

The considered benchmark schemes are listed as follows: 1) \textbf{FPA}: The $N$-dimensional FPA-based ULA with half-wavelength antenna spacing is considered. The optimal AWV is obtained according to Section III-D. 2) \textbf{Alternating position selection (APS)}: The 1D line segment for MA positioning is quantized into discrete locations with equal-distance $D_0=\lambda/2$. We apply Algorithm~\ref{alg1}, where steps 10 and 11 for the optimization of $\bm{x}$ with a given $\bm{w}$ are replaced by  sequentially optimizing one MA's position with the other $(N-1)$ MAs' positions being fixed, and an exhaustive search  over discrete locations is utilized for each MA. 3) \textbf{Algorithm~\ref{alg1} without iteration (AWI)}: Problems (P2) and (P3) are solved once without outer iterations in Algorithm~\ref{alg1}.

\begin{figure}[!t]
	\centering
	\includegraphics[width=75mm]{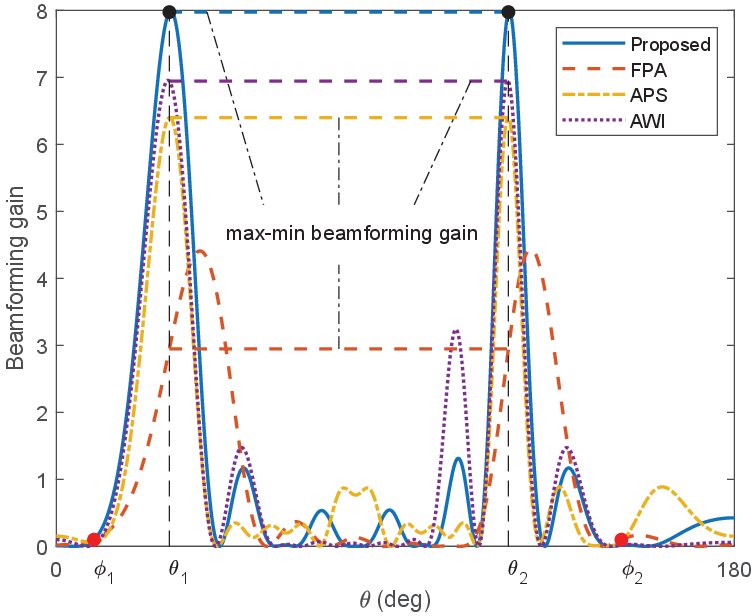}
	\caption{Comparison of beam patterns with MA and FPA arrays.}
	\label{beam_pattern}
\end{figure}

In Fig.~\ref{beam_pattern}, we illustrate the beamforming patterns for the proposed method and the benchmark schemes. We set $D=8\lambda$, $N=8$, $K=2$, $L=2$, $\theta_1=30^{\circ}$, $\theta_2=120^{\circ}$, $\phi_1=10^{\circ}$, and $\phi_2=150^{\circ}$. From Fig.~\ref{beam_pattern}, the proposed method achieves much higher max-min beamforming gain than the benchmark schemes and even approaches close to the full beamforming gain (i.e., $N = 8$). The proposed method can achieve $99.7\%$ of the full beamforming gain, while the FPA, APS, and AWI schemes achieve $36.9\%$, $80.0\%$, and $86.8\%$ of it, respectively. This is because as compared to the benchmark schemes, the proposed method can adjust the steering vector $\bm{\alpha}(\bm{x},\theta)$ by optimally exploiting the new degrees of freedom (DoFs) in the APV $\bm{x}$, such that the correlations among all signal directions $\{\bm{\alpha}(\bm{x},\theta_k)\}_{k=1}^K$ are maximized, while the correlation between each pair of signal direction in $\{\bm{\alpha}(\bm{x},\theta_k)\}_{k=1}^K$ and interference direction in $\{\bm{\alpha}(\bm{x},\phi_l)\}_{l=1}^L$ is minimized.

\begin{figure}[!t]
	\centering
	\includegraphics[width=75mm]{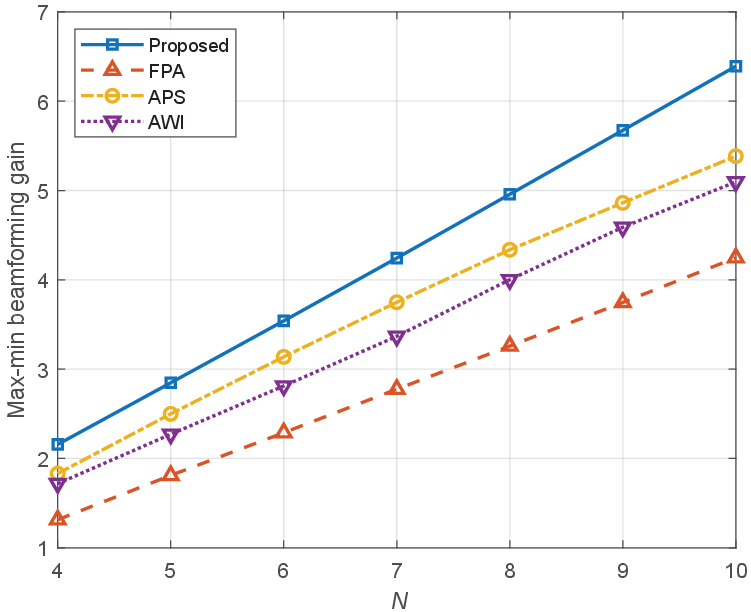}
	\caption{Max-min beamforming gain versus $N$.}
	\label{N}
\end{figure}

\begin{figure}[!t]
	\centering
	\includegraphics[width=75mm]{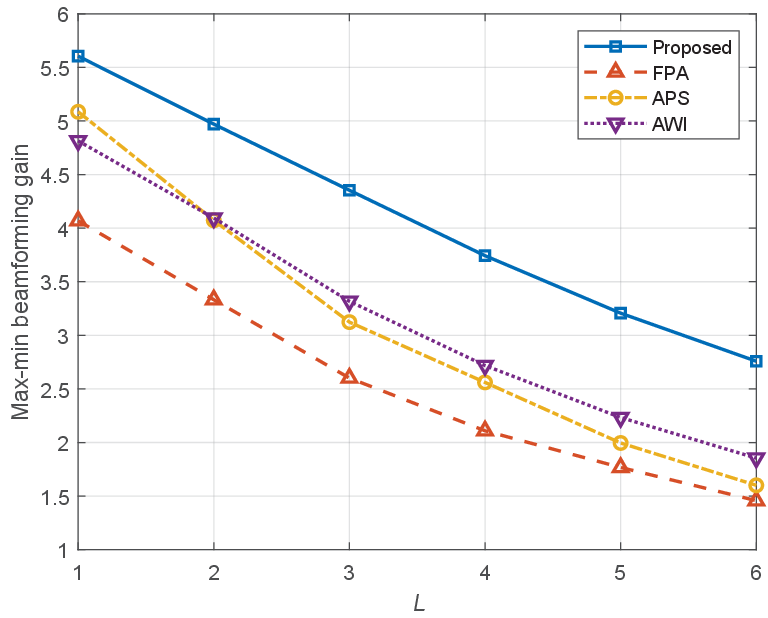}
	\caption{Max-min beamforming gain versus $L$.}
	\label{L}
\end{figure}

Fig.~\ref{N} compares the max-min beamforming gain versus $N$ for different schemes. We set $D=10\lambda$, $K=2$, and $L=2$. $\{\theta_k\}_{k=1}^K$ and $\{\phi_l\}_{l=1}^L$ are randomly generated following uniform distribution over $[0, 180^{\circ}]$. From Fig.~\ref{N}, the proposed method always outperforms all benchmark schemes, and the performance gap increases with $N$. Moreover, the proposed method still has $64.1\%$ performance improvement over the FPA scheme with only $N=4$ MAs' positions that can be optimized. This result shows that the proposed method can be applied to practical multiple-MA systems even with small numbers of MAs.

In Fig.~\ref{L}, we compare the max-min beamforming gain versus $L$ for all schemes. We set $D=8\lambda$, $N=8$, and $K=2$. It is observed again that the proposed method always outperforms all other schemes, demonstrating that the proposed method is robust to the number of interference directions in practice. Even under a large number of interference directions (e.g., $L=6$), the proposed multi-beam forming method with MA arrays can still achieve large performance improvement (about 3 dB gain) as compared to that with the FPA arrays.

\section{Conclusions}
In this letter, we studied the multi-beam forming with MA arrays by exploiting the new DoFs via antennas' position optimization. We aimed to maximize the beamforming gain over all signal directions subject to a given constraint on the maximum power over all interference directions, via the joint optimization of the APV and AWV. An alternating optimization algorithm was proposed to obtain a suboptimal solution by iteratively optimizing one of the APV and AWV with the other being fixed. Simulation results showed that our proposed multi-beam forming design with MA arrays can significantly enhance the max-min beamforming gain with effective interference suppression compared to that with the conventional FPA arrays.

\bibliographystyle{IEEEtran}
\bibliography{IEEEabrv,IEEEexample}

\end{document}